\documentclass[aps,prb,twocolumn]{revtex4-1}

\usepackage{graphicx}
\usepackage{amssymb}
\usepackage{amsfonts}
\usepackage{amsmath}
\usepackage{amsbsy}

\renewcommand{\vec}[1]{\mbox{\boldmath$\mathrm{#1}$}}

\begin{document}

\title{Magnetic dynamics driven by the spin-current generated via  spin-Seebeck effect}
\author{Chenglong Jia$^{1,2}$ and Jamal Berakdar$^{1}$}
\affiliation{1. Institut f\"ur Physik, Martin-Luther Universit\"at Halle-Wittenberg, 06099  Halle, Germany\\
2. Key Laboratory for Magnetism and Magnetic Materials of the Ministry of Education, Lanzhou University, Lanzhou 730000, China}

\begin{abstract}
We consider the spin-current driven  dynamics of a magnetic nanostructure
in a conductive magnetic wire under a heat gradient in an open circuit, spin Seebeck effect geometry. It is shown that the spin-current
scattering results in a spin-current  torque  acting on the nanostructure  and leading to precession and  displacement. The scattering leads also to a redistribution of the spin electrochemical potential along the wire resulting in a break of the polarity-reversal symmetry of the
inverse spin Hall effect voltage with respect to the heat gradient inversion.
\end{abstract}


\maketitle

\begin{figure}[b]
  \centering
  \includegraphics[width=0.45\textwidth]{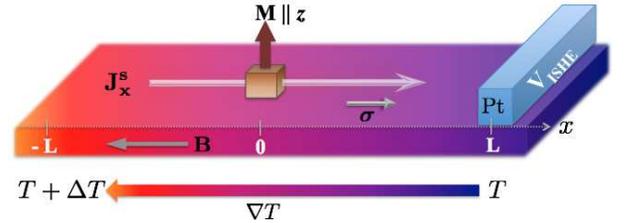}
  \caption{A localized magnetic structure $\vec M$  in a ferromagnetic conductor of length $2L$
   subject to the constant temperature gradient $\vec \nabla T$.
 $\vec B$   is a  saturation magnetic field along  $\vec \nabla T$.  The pure spin current  $\vec{J}^s_x$ is
 signaled by the inverse spin Hall voltage $V_{\mbox{ISHE}}$  measured  in a Pt strip. $x$ and $z$ axes are indicted.}
  \label{fig:system}
\end{figure}

\emph{Introduction.- }The discovery in the 1820s by T.J. Seebeck that due to a temperature gradient an electric voltage emerges  along the temperature drop, revealed
the relationship between heat and charge currents.
The reversal of Seebeck's effect, i.e. the appearance of a temperature gradient upon an applied voltage, was shortly thereafter confirmed by J.-C. Peltier in 1834. In addition to other  thermo-electric phenomena such as the Joule heating, in magnetic fields new thermo-magnetic effects arise:   A resistive conductor with a  temperature gradient $\vec\nabla T$  placed in a  magnetic field $\mathbf B$ perpedicular to  $\vec\nabla T$
develops a potential drop  normal to both  $\vec\nabla T$ and $\mathbf B$. This phenomenon is termed  the Ettingshausen effect and its reverse is the Nernst effect.   In a magnetic material the anomalous Nernst effect occurs (i.e.\ the Nernst effect due to the spontaneous magnetization) which was
  first observed  for Ni and Ni-Cu alloy \cite{1,2}. Recently, in Refs.\cite{3,4,5,6}  measurements of the planar and the anomalous Nernst effect were reported for a variety of materials including magnetic semiconductor, ferromagnetic metals, pure transition metals, oxides, and chalcogenides.
A qualitatively new phenomena, the Spin-Seebeck effect (SSE), was discovered  2008 by Uchida \emph{et al.} \cite{7} showing that in a ferromagnetic material (a mm-size Ni$_{81}$Fe$_{19}$ sample) and in an open-circuit geometry (which is also the geometry studied in this work, cf. Fig. \ref{fig:system})
a heat current results in a spin current, i.e. a flow of spin angular momentum and hence a spin voltage, even if $\vec\nabla T$ is parallel to $\mathbf B$ (where the Nernst-Ettingshausen effect does not contribute).  The spin voltage is reflected by
 a charge voltage  that emerges (due the inverse spin Hall effect (ISHE)) in a Pt strip  deposited on the sample perpendicular to $\vec\nabla T$ (cf. Fig.\ref{fig:system}).  Further experiments on resistive conductors (\cite{8} for Ni$_{81}$Fe$_{19}$),  insulating ferrimagnets (LaY$_{2}$Fe$_{5}$O$_{12}$ in \cite{9}), and for ferromagnetic semiconductors (GaMnAs \cite{10}) underline the generality of SSE.  These fascinating effects
 are not only of fundamental importance; thermo-electric elements are already indispensable  for temperature sensing and control and for
  current-heat conversion. SSE opens the way for  thermo-electric spintronic devices with  qualitatively  new tools for energy-consumption reduction. It is highly desirable to explore whether SSE can be utilized to
  steer localized magnetic textures, as problem addressed here.
  Theoretically, the reciprocity between the dynamics in the magnetic order and the heat gradients is governed by the Onsager relations.
  The Onsager matrix were discussed from a general point of view in Ref. \cite{12}  with a focus on the transport of charge, magnetization, and  heat.
%
  In Refs.\cite{13,14} a thermo-magnetic mesoscopic circuit theory was put forward.  Ref. \cite{15} pointed out the occurrence of
   a thermally excited spin current in  resistive conductors with an embedded ferromagnetic nanoclusters. Other recent works \cite{13,14}
   addressed the thermally induced spin-transfer torque in a  spin valves structures whereas the phenomenological study \cite{16} is focused on the spin-transfer torques in quasi one-dimensional magnetic domain walls (DWs) by introducing a viscous
    term into the Landau-Lifshitz-Gilbert equation (LLG). \\
  \emph{Spin current-}  The microscopic mechanism for the appearance of the spin current in SSE is not yet completely understood, it is however
   an experimental fact that  in the geometry of Fig.\ref{fig:system} the thermal gradient $\vec{\nabla} T$ generates a steady state spin current $\vec{J}^s_T$  without a charge current \cite{7,8,9,10}. The amplitude of  $\vec{J}^s_T$ is found to be determined by \cite{7,8,9,10}
   \begin{equation}
  \vec{J}^s_T = - \kappa  \vec{\nabla} T
\label{eq:js}\end{equation}
where $\kappa$ is a temperature-independent transport coefficient whose properties are discussed in \cite{7,8,9,10}; no charge current is generated.
The purpose of this work is to inspect the dynamics triggered by $\vec{J}^s_T$ (eq.(\ref{eq:js})) for the case where localized magnetic texture $\vec M$  \cite{note-1} is present in the  ferromagnetic (FM) conductor (cf. Fig.\ref{fig:system}), a problem of great importance and has not been addressed so far.
As show below,  the quantum-mechanical scattering of $\vec{J}^s_T$ from $\vec M$
  acts in effect with a spin-current  torque  on $\vec M$  which results in an oscillatory and a displacement motion of $\vec M$.
   Upon scattering $\vec{J}^s_T$ also changes. This leads to   a redistribution of the spin electrochemical potential which can be measured via ISHE.

 The system under consideration is illustrated in  Fig. \ref{fig:system}.  Two thermal reservoirs with different temperatures $T$ and $T + \Delta T$ create along the $x$ axis in a FM conductive wire of length $2L$ a steady $T$-gradient $\vec{\nabla} T$ and hence a steady-state spin current $\vec{J}^s_x$.  This means, without knowing the detail of the operators associated with SSE, these project the system onto a chargeless eigenstate $\psi(x)$ of the spin current operator $\vec{j}_{\mu}^s (k)$.
%
  Generally,  such a state can be written as \cite{note-2}
\begin{equation}
   \psi(x) =\frac{1}{2} \left[e^{ikx} \binom{e^{i\phi}}{e^{-i \phi}} + e^{-ikx}\binom{e^{i\theta}}{e^{-i\theta}}\right].
\label{eq:psi}\end{equation}
The $x$ expectation value of the charge current $ \vec{j}^e (k) = \frac{i\hbar}{2m} [ (\vec{\nabla} \psi^{\dag} (x)) \psi(x) - \psi^{\dag}(x) \vec{\nabla} \psi (x) ]$ vanishes, i.e. $j^e_x (k) \equiv 0$ (here $m$ is the effective mass).
In contrast, for the spin current $ \vec{j}_{\mu}^s (k) = \frac{i\hbar}{2m} [ (\vec{\nabla} \psi^{\dag} (x))  \sigma_{\mu} \psi(x) - \psi^{\dag}(x) \sigma_{\mu} \vec{\nabla} \psi (x)]$ we infer
\begin{eqnarray}
& j^s_x(k) = \frac{\hbar k}{2m} (\cos 2 \phi -\cos2 \theta), \\
& j^s_y(k) = \frac{\hbar k}{2m} (\sin 2 \theta -\sin 2\phi).
\end{eqnarray}
In general, the thermal transport is ballistic \cite{T-transport} but with diffusive spins, i.e., upon creating (\ref{eq:psi})
  the spin coherence is lifted
 by scattering events that randomize   $\phi$ and $\theta$ within $[0, 2\pi]$. Hence,
   the expectation value of the spin current vanishes on the scale of the spin-flip diffusion length \cite{semicl-theory,spin-pumping}, \emph{i.e.}, $\vec{J}^s_{\mu}(k) = \oint  \vec{j}^s_{\mu} d\theta d\phi /(2\pi)^2  =0$. However, when the wire is magnetically polarized and driven to  saturation by the magnetic field  $\vec B$ \cite{7,8,9,10}, we find $\langle \sigma_x \rangle \neq 0$ but $\langle \sigma_y \rangle = 0$. Eq.(\ref{eq:psi})  reads then for an exchange-split conductor
\begin{equation}
   \psi_{B} (x) =\frac{1}{2} \left[ e^{ikx} \binom{1}{1} + e^{-ikx} \binom{e^{i\theta}}{e^{-i\theta}} \right].
\label{eq:psiB}\end{equation}
Here $\theta \in [0, 2\pi]$ still appears due to the residual spin precession and diffusion.  Then we have $J^e  \equiv 0$,  $J^s_y(k) = \oint j^s_y d\theta/2\pi = 0 $, whereas $J^s_x(k) = J^s_0(k) = \hbar k/2m$ in line with the experimental observation \cite{7,8,9,10}.

\begin{figure}[b]
  \centering
  \includegraphics[width=0.4\textwidth]{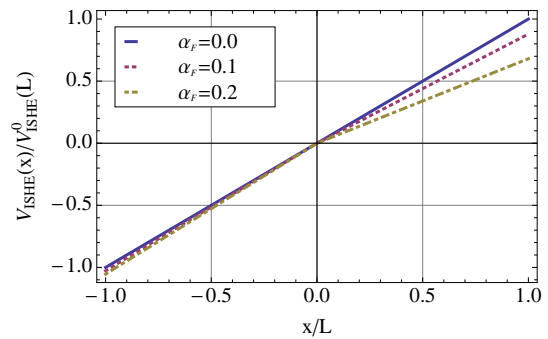}
  \caption{The electric voltage $\text{V}_{\text{ISHE}}$ generated by the inverse spin-Hall effect in the Pt layer as a function of the magnetic scattering strength $\alpha_F = gM_{0}m/k_{F}\hbar^2$ with $k_F$ being the Fermi wave vector. The relaxation rate is $\Gamma/E_F =0.01$. }
  \label{fig:Vh}
\end{figure}

The main purpose of the present work is to investigate the influence of a localized magnetic, non-diffusive 
scatterer $\vec M$ (where $x=0$ is taken as its central position  (see Fig.\ref{fig:system})).
$\vec M$  has a uniaxial anisotropy along an axis chosen to be $z$. If  $\vec M(x)$ has an internal structure, e.g.\ a non-collinearity,  which varies
on a scale larger than $\lambda=2\pi/k$ (the variation scale of (\ref{eq:psiB})), one can unitary transform to align locally with $\vec M$ which introduces
a weak gauge potential that can be dealt \cite{tobepub} with in a perturbative  way using the Green's function constructed from (\ref{eq:psiB}) (similarly as done in \cite{nick,jpa}). We find $\vec M$ has a stronger influence if its range of variation is comparable to $\lambda$.
 In this case the magnetic texture acts in effect as  $\vec{M}(x) = \vec{M}_0 \delta (x)$, where the magnetic moment $\vec{M}_0$ derives
  from an average of
$\vec{M}(x)$ over its extension $w$. The model is realizable for \cite{10} rather then for metals. 
 The interaction between $\vec{M}$ and the electron spin $\vec{\sigma}$ reads \cite{M-nanowire},
\begin{equation}
   H_{int} = g \vec{M}(x) \cdot \vec{\sigma}
\label{eq:hint} \end{equation}
where $g$ is a local coupling constant and $M_0$ is large enough to be treated classically. For $\vec{M}_0$ aligned with $z$ axis, as in Fig.\ref{fig:system}, we derive
 using  eqs.(\ref{eq:psiB},\ref{eq:hint}) the expression for the spinor wave function in the presence of $\vec M(x)$, namely
\begin{equation}
\psi_s (x) =
   \begin{cases}
      \frac{e^{ikx}}{2} \binom{1}{1} + \frac{e^{-ikx}}{2} \binom{r}{r^{\star}} + \frac{e^{-ikx}}{2}\binom{t e^{i\theta}}{t^{\star} e^{-i\theta}} &  \text{for } x < 0, \\
      \frac{e^{-ikx}}{2}\binom{e^{i\theta}}{e^{-i\theta}} +\frac{e^{ikx}}{2}\binom{re^{i\theta}}{r^{\star}e^{-i\theta}} +  \frac{e^{ikx}}{2} \binom{t}{t^{\star}} & \text{for } x > 0.
    \end{cases}
  \label{spinor}
\end{equation}
The scattering state $\psi_s (x)$ describes the spinor wave function in the original spin channel, which is partially reflected into the original-spin and the spin-flip channels, and also partially transmitted into theses two channel, which gives the complex spin reflection and  transmission coefficients $r$ and $t$,
\begin{equation}
  r = - \frac{i \alpha}{1+ i \alpha}, \;  t = \frac{1}{1+i \alpha},\quad \alpha = g M_0 m/k\hbar^2.
\label{eq:a}\end{equation}
The magnetic scattering gives rise to a short pseudo-circuit to the charge channels, for we find
\begin{equation}
  j^e_x = \frac{2 \alpha  \sin \theta }{1+\alpha ^2}.
\end{equation}
 $\langle j^e_x \rangle$ vanishes however beyond the spin diffusion length after averaging over $\theta$.
 Counterparts, i.e. a  pure spin current  generated by  a charge current  when scattered off a magnetic structure,
  are well known, e.g. \cite{M-nanowire,MTJ,miguel}.
  The spin current carried by (\ref{eq:psiB}) is  modified upon scattering and   a none-zero $J^s_y$ emerges
\begin{eqnarray}
 && J_x^s/ J^s_0=
           \begin{cases}
              \frac{1+3 \alpha ^2}{(1+\alpha ^2 )^2} & \text{for } x < 0, \\
              \frac{1-\alpha ^2}{(1+\alpha ^2 )^2}    & \text{for } x >0,
           \end{cases} \\
 && J_y^s / J^s_0 =
           \begin{cases}
              \frac{2 \alpha ^3}{(1+\alpha ^2)^2}  & \text{for } x < 0, \\
              \frac{2 \alpha }{ (1+\alpha ^2)^2}     & \text{for } x < 0.
           \end{cases}
\label{Js}
\end{eqnarray}
%
 Within linear response, the spin voltage along the wire is
$\mu_{s} (x) = \xi_s x \langle J^s_x (\alpha) \rangle $ where $\xi_s$ is a function of the elementary charge, the spin-dependent electric conductivity, the spin-dependent Seebeck coefficient, and the spin-Seebeck coefficient of the FM wire \cite{phe-theory}. The
 quantum-mechanically averaged spin current is
$\langle J^s_x (\alpha)\rangle = Tr( J^s_x (\alpha) \rho )$ \cite{dbbprb}. The single electron density matrix is $\rho = -\frac{1}{\pi} \text{Im} \sum_{\vec{k}} \frac{\psi_s \psi_s^{\dag}}{E_F - E_{\vec{k}}+ i\Gamma}$ where $E_F$ is the Fermi energy, $E_{\vec{k}} = \hbar^2 k^2/2m + \hbar^2 \vec{k}_{\|}^2/2m$ with $\vec{k}_{\|}$ being the transverse wave vector, and $\Gamma$ is a Lorenzian relaxation rate due to  disorder \cite{LorenzianWidth}. Depositing
a conductive strip with a strong spin-orbit coupling, e.g. Pt, as shown
 in Fig.\ref{fig:system}, $\mu_{s} (x)$ can be imaged via
 the electric voltage $V_{\text{ISHE}}$ generated by the inverse spin-Hall effect in Pt using the relation
\begin{equation}
 \frac{ \text{V}_{\text{ISHE}} (x)}{\text{V}^0_{\text{ISHE}}(L)} = \frac{\langle J^s_x (\alpha) \rangle }{\langle J^s_x (0) \rangle } \frac{x}{L}
\end{equation}
where $\text{V}^0_{\text{ISHE}} (x)$ is the electric voltage measured in Pt in absence of the magnetic scatterer $\vec M$. Explicitly,  $\text{V}^0_{\text{ISHE}} (x)= \gamma \xi_s x \langle J^s_x (0) \rangle$ with $\gamma$ being a system-dependent parameter \cite{7,8,9,10}, determined by the spin-Hall angle in Pt, the spin-injection efficiency across the FM/Pt interface, and the length and the thickness of Pt wire.
Due to the spin current scattering off $\vec M$
the Hall voltage $\text{V}_{\text{ISHE}}$ losses its odd  symmetry with  respect to a reflection
 at $x=0$, i.e.  we deduce $-\text{V}_{\text{ISHE}}(-x) \neq \text{V}_{\text{ISHE}} (x)$.  As shown in Fig.\ref{fig:Vh}
 the amount of the symmetry break depends on $\alpha$, and can be taken in the experiment as an indicator of the
 presence of magnetic scattering centers.

\emph{Magnetization dynamics.}-  In as much as $J^s_x $ is modified
by the presence of $\vec M$, the scattering triggers a dynamics of $\vec M$ which is usually much slower than the carrier scattering
 dynamics and can be classically treated ($M_0$ is assumed large).
$J^s_x $ acts on $\vec M$ with  a torque  $T_{\mu}$ that follows
from the jump in the spin current at the point $x=0$, $T_{\mu} = J^s_{\mu} (0^-) - J^s_{\mu} (0^+)$.
Hence, $T_{\mu}$  derives from our quantum mechanical calculations as
\begin{eqnarray}
  T_x  = J^s_0 \frac{4 \alpha^2}{(1+\alpha^2 )^2}, \; T_y  = - J^s_0 \frac{2 \alpha (1-\alpha^2 )}{(1+\alpha^2)^2}.
  \label{storque}
\end{eqnarray}
 Both components are transversal.  $T_y$ tends to align $\vec M$ to the direction of the FM magnetization, while $T_x$ tries to rotate the moment  $\vec M$ around the axis $\hat{e}_x$.
Equivalently, within our model, the  spin-current torque $T_{\mu}$ is obtained from the  spin density  $S_{\mu} (x)$ accumulated at the localized moment (due to  interference of
incoming and reflected waves) as
\begin{equation}
   T_{\mu} = - \frac{gM_0}{\hbar} [\vec{n} \times \vec{S}(x=0)]
\end{equation}
where $\vec{n}$ is the unit vector along $\vec{M}$, and  the  spin density we obtain from $S_{\mu} (x) = \psi^{\dag}(x) \sigma_{\mu} \psi(x)$.
%
%
Since $M_0$ is assumed large (say $\geq 5/2\; \mu_B$) the spin current-induced magnetization dynamics
 can be treated with the  modified  LLG equation \cite{emLLG,smLLG}
\begin{equation}
  \frac{\partial \vec{n}}{\partial t} = \frac{D_z}{\hbar} [\vec{n} \times \hat{e}_z] + \frac{a_g}{\hbar} \vec{n} \times \frac{\partial \vec{n}}{\partial t} -\frac{g}{\hbar} [\vec{n} \times \vec{S}(0)]
  \label{LLG}
\end{equation}
where $D_z$ is the anisotropy energy and $a_g$ is the Gilbert damping parameter \cite{footn2}.  Two motion types  of $\vec M$ occur:
%
\begin{figure}[t]
  \centering
  \includegraphics[width=0.5\textwidth]{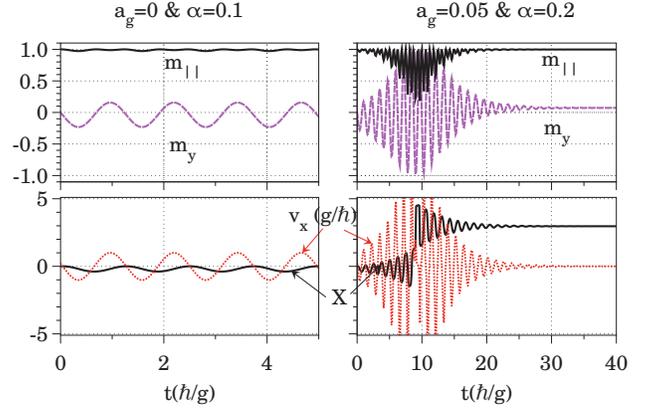}
  \caption{Precession of the magnetic moment $\vec{M}$ (eq.(\ref{eq:n1})) for different
   spin-current scattering strength $\alpha$ (cf. eq.\ref{eq:a}) and Gilbert damping $a_g$. Here $v_x = \partial X/ \partial t$
   and
   $D_z/g =5$. }
  \label{fig:oscillation}
\end{figure}
%
\begin{figure}[t]
  \centering
  \includegraphics[width=0.5\textwidth]{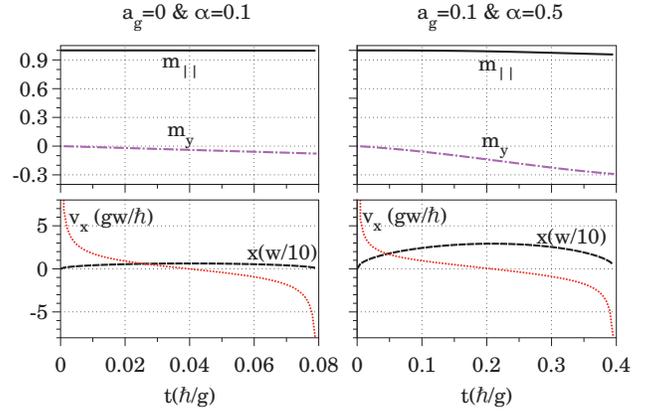}
  \caption{Displacement of the magnetic moment (given by eq.(\ref{eq:n2}))  vs. time
  for the same parameters of Fig.\ref{fig:oscillation}. $v_x = \partial x/ \partial t$.}
  \label{fig:displacement}
\end{figure}

\emph{Precession-} Introducing the following magnetization distribution

\begin{equation}
   \vec{n} = [m_{\|}(t) \sin X(t), m_y(t), m_{\|}(t) \cos X(t)]
\label{eq:n1}\end{equation}
and propagating with the LLG equation (Eq.\ref{LLG}) starting from   $ m_{\|}(0) =1,\; m_{y}(0) =0,\; \text{and } X(0) =0$,
 we calculate the time dependence  of $\vec M$ shown in Fig.\ref{fig:oscillation}.
The oscillations of $X(t)$ results in small $x$ and $y$ components of the magnetization.  The magnetic moment precesses with a velocity $v_x (t)= \partial X(t) /\partial t$ in the presence of the SSE-generated spin current  ($v_x(0) \neq 0$).  We note that the maximum deflection angle $X_{max}$ depends implicitly on the spin current  dynamics through the parameter $\alpha$, as determined by eq. (\ref{eq:a}). Also the Fermi energy enters through the $k$ dependence of $\alpha$. The dynamics is
a mixture of anisotropy-dominated precession and damping.

\emph{Displacement-} Let us consider the initial localized magnetic moment distribution
\begin{eqnarray}
   \vec{n} &=& [m_{\|}(t) \sin \zeta , m_y(t), m_{\|}(t) \cos \zeta ]\nonumber\\
%
%
  \zeta &=& \cos^{-1} \left[ \tanh^2(x/w) \right]
\label{eq:n2}\end{eqnarray}
where $w$ stands for the extension of the localized moment and $m_{\|}(0) =1,\; m_{y}(0) =0,\; x(0) =0.$
 As concluded from  Fig.\ref{fig:displacement}, the moment is set in motion when subjected to the spin current. The velocity changes from positive to negative, which is different from the motion of a single N\'eel wall \cite{DWs} (the velocity decreases to zero in a fraction of a nanosecond, and the DWs stops completely.).

\emph{Remarks and conclusions.-}
 Our main  result is that the SSE generated spin current in a wire may scatter from a localized magnetic structure setting it
 in  an precessional and a displacement motion. The scattering also leads to a redistribution of the spin current in the wire and hence changes
 the ISHE signal. From these results conclusion can be made on the influence of a collection of non-interacting localized moments but no
 statement can be made when they interact or even form clusters. We also note that the present conclusions do not apply to a domain wall (DW),
 (except for very close non-resonant (transversal) $180^\circ$ DW pair, e.g. as  in \cite{vitaprl}). In fact, our initial finding \cite{tobepub}
 is that a single sharp DW is  less affected by the spin current because the spin-current torques acting from left and right of the DW partially compensate. This is not so for an adiabatic or asymmetric DW because the $T$-gradient modifies the DW along $\vec\nabla T$. As for the experimental observation
 of the magnetic moment dynamics,
 it should be noted that the temperature gradient has to be sustained on the time-scale of this dynamics; a fast (e.g., femtosecond)  strong heat pulse is inappropriate for our (constant $\nabla T$, linear response) study and may cause locally a longitudinal dynamics of $\vec M$.


\begin{thebibliography}{99}
\bibitem{note-1} As an example, we take a single-molecule magnets Mn$_{12}$ [Mn$_{12}$O$_{12}$(O$_2$C-C$_6$H$_4$-SAc)$_{16}$(H$_2$ O)$_4$] as the local magnetic scatterer $\mathbf{M}$.  The total spin and diameter of Mn$_{12}$ are $M_0 =10$ and $w=3$nm, respectively \cite{Mn12}. The local exchange coupling between the carriers and the moment $\mathbf{M}$ is given as $gM_0 =1$meV \cite{SMMT}.
\bibitem{note-2} We employ a continuous model; a discretized  treatment based on  Heisenberg spins is straightforward and does not alter qualitatively the  amplitude of the spin/charge current.
\bibitem{footn2} Temperature effects
can be included in eq.(\ref{LLG}) as in \cite{alexander}. by assuming $\vec M$ to be in a local thermal equilibrium at the
temperature $T(x=0)$ (assuming $\vec\nabla T = constant$) an applying the fluctuation-dissipation theorem which yields
a stochastic field that adds to the effective field in eq.(\ref{LLG}). In \cite{alexander} it is shown that at low $T$ thermal fluctuations
do not alter qualitatively the LLG dynamics.
\end{thebibliography}
\end{document}